\documentclass{amsart}

\usepackage{amsfonts}
\usepackage{amsmath}
\usepackage{amssymb}

\pdfoutput=1

\usepackage[pdftex]{graphicx}

\begin{document}

\newcommand{\Rzp}{{\mathbb{R}_{\geq 0}}}
\newcommand{\oh}{{\scriptstyle\frac{1}{2}}}
\newcommand{\txtfrac}[2]{{\scriptstyle\frac{#1}{#2}}}

\newcommand{\eolm}{\hspace*{\fill}\mbox{$\Box$}}
\newcommand{\eqeolm}{\hspace*{\fill}\Box}

\title[Metric distances derived from correlations]{Metric distances derived from cosine similarity and Pearson and Spearman correlations}
\author{Stijn van Dongen and Anton J. Enright}
\address{EMBL-EBI, Hinxton, Cambridge, UK}
\email{stijn@ebi.ac.uk}

\begin{abstract}
      We investigate two classes of transformations of cosine similarity and
      Pearson and Spearman correlations into metric distances, utilising the
      simple tool of metric-preserving functions.  The first class puts
      anti-correlated objects maximally far apart.  Previously known
      transforms fall within this class.  The second class collates correlated
      and anti-correlated objects.  An example of such a transformation that
      yields a metric distance is the sine function when applied to centered data.
\end{abstract}

\maketitle

\section{Results}
   We derive metric distances from the sample Pearson coefficient, sample Spearman coefficient,
   and cosine similarity.
   Using~$A$ to denote any of these, it is already known that
   $\theta = \arccos(A(x,y))$ yields a metric distance, known as the angular distance.
   We further obtain the correlation distance~$\sin(\oh\theta)$,
   or equivalently $\sqrt{\oh(1-A(x,y))}$.
   Both distances place anti-correlated objects maximally far apart.
   A second class of metric distances is obtained that collate correlated and anti-correlated
   objects. Examples are the acute angular distance~$\oh\pi - |\oh\pi - \theta|$
   and the absolute correlation distance~$\sin(\theta)$, or equivalently~$\sqrt{1-A(x,y)^2}$.

\section{Background}
   The Pearson correlation coefficient, Spearman correlation coefficient and
   the cosine similarity are staples of data analysis.  The Pearson and
   Spearman coefficients measure strength of association between two
   variables~$X$ and~$Y$. The Pearson coefficient, commonly denoted by~$\rho$,
   is defined as the covariance of the two variables divided by the product of
   their respective standard deviations.
\begin{equation}
\label{eq::pearson}
\rho_{X,Y} = \frac{{\mathrm cov}(X,Y)}{\sigma_X \sigma_Y}
\end{equation}
   The Spearman coefficient is obtained by applying the Pearson coefficient to
   rank-transformed data.  Both are unaffected by linear transformations of the
   data.
   Given vectors~$x$ and~$y$, respectively sampling~$X$ and~$Y$ and each of
   length~$n$, the sample Pearson coefficient~$r_{x,y}$ is obtained by
   estimating the population covariance and standard deviations from the samples, as
   defined in Equation~(\ref{eq::sample:pearson}). Here
   $\overline{x}$~and~$\overline{y}$ denote the sample means.
\begin{equation}
\label{eq::sample:pearson}
r_{x,y} = \frac{\sum (x_i - \overline{x}) (y_i - \overline{y})}{\sqrt{\sum (x_i - \overline{x})^2} \sqrt{\sum (y_i - \overline{y})^2}} 
\end{equation}
   The cosine similarity is a standard measure used in information retrieval.
   It is the cosine of the angle between two Euclidean vectors,
   and thus unaffected by scalar transformations in the data.
   It is defined below in Equation~(\ref{eq::cosine}) for vectors~$x$ and~$y$.
\begin{equation}
\label{eq::cosine}
\frac{\sum x_i y_i}{\sqrt{\sum {x_i}^2}\sqrt{\sum {y_i}^2}}
\end{equation}
   These measures are related;
   Pearson is identical to the cosine applied to centered data (centered
   cosine), as evident from equations~(\ref{eq::sample:pearson}) and~(\ref{eq::cosine}).
   For the purpose of this paper the terminology of \textit{vectors}
   and \textit{samples} is used interchangeably. We are not concerned
   with statistical properties of the Pearson coefficient under certain models,
   but solely interested in its properties as a function mapping Euclidean spaces
   to the interval~$[-1,1]$.
   We will henceforth refer to Pearson, Spearman, and cosine similarity as $P$, $S$,
   and $C$, and use $A$ to indicate all of them are applicable.

   Where dissimilarities are used, it is desirable that they satisfy the
   triangular inequality and are thus a metric distance.  Informally, this
   means that detours take longer: the distance from $a$ to $c$ should always be
   shorter than the distance from $a$ to $b$ plus the distance from $b$ to $c$.
   Metric distances abound in data analysis, formalizing
   a property that is intuitively expected and that allows stringent reasoning
   about data points.  Several methods require this, such as for building
   $M$-trees \cite{Ciaccia97m-tree:an}
   and accelerated algorithms that use the triangle inequality to skip
   computations by tracking bounds \cite{Brin:1995:NNS:645921.673006, hamerly:making}.


\section{Metric distances}
   A metric distance takes as input two objects and outputs a real number.
   It requires four properties. These are i) all distances are
   nonnegative, ii) the distance of an object to itself is zero and distinct
   objects are never at distance zero, iii) the distance between two objects is
   the same in both directions, and iv) the distance satisfies the property that
   detours are longer, more commonly stated as the triangle inequality. More
   formally, given a distance $d$, it states that $d(x,y) \leq d(x,z) + d(z,y)$ for all
   objects $x$, $y$, and $z$. In this formulation, the distance between $x$ and $y$
   is compared to the distance when using $z$ as a detour.

   In the analysis of distances derived from correlations and cosine similarity
   we will use a class of functions called \textit{metric preserving}.
   A function~$f$ is metric preserving if the
   distance~$d_f(x,y) = f(d(x,y))$ is again metric for any metric~$d$.
   More specifically, we shall make use of an important subclass of metric-preserving
   functions, namely those that are \textit{concave} and increasing.
   A function $f$ is called concave on an interval $I$ if for all $x$ and $y$
   in $I$ and for $t$ in $[0,1]$ the inequality
\begin{equation}
\label{eq::chord}
   f(tx+(1-t)y) \geq t f(x) + (1-t)f(y)
\end{equation}
   holds. We refer to this as the chord condition.
   It is the formal way of stating that the chord drawn from $[x,f(x)]$ to
   $[y,f(y)]$ does not exceed $f$ in the interval $[x,y]$.  It essentially
   means that $f$~is curving inward on~$I$, as shown in the figure below.\\[2ex]
\hspace*{1cm}
\includegraphics{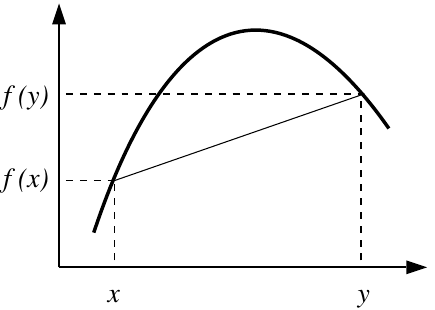}

   The following lemma, relating \textit{concave} functions to metric
   preserving functions is well-known (see e.g.  \cite{mpf::corrazza}). We
   include a short detailed proof as it is an important prerequisite to this
   paper, consisting of several steps gathered here for ease of reference.  It
   shows subadditivity to be the key property making certain concave functions
   also metric-preserving.

\textbf{Lemma 1}
   For~$f$ to be metric preserving it is sufficient if~$f(0)=0$,
   and $f(x)$~is both \textit{increasing} and \textit{concave} for $x>0$.

\textbf{Proof}
   We first prove that functions that are concave for $x>0$ and
   satisfy $f(0) \geq 0$ are also
   \textit{subadditive} for~$x\geq 0$ (that is, $f(a+b)\leq f(a)+f(b)$
   for $a,b \geq 0$).
   This follows by setting $y=0$ in the chord condition~(\ref{eq::chord}) and
   using the postulate $f(0) \geq 0$. We obtain
   the scalar inequality $t f(x) \leq f(t x)$, for $0
   \leq t \leq 1$.
   We then rewrite $f(a+b)$ as
   $\txtfrac{a}{a+b} f(a+b) + \txtfrac{b}{a+b} f(a+b)$, noting that
   $\txtfrac{a}{a+tb}$ and  $\txtfrac{b}{a+tb}$ both lie in $[0,1]$.
   Using the scalar inequality just derived we bound the rewritten expression
   from above by $f(\txtfrac{a}{a+b}(a+b)) + f(\txtfrac{b}{a+b}(a+b))$,
   equaling $f(a)+f(b)$.

\enlargethispage{2\baselineskip}
   The proof of the lemma can now be concluded. We need to prove
   that $d_f$ is a metric distance, i.e. $d_f(x,y) \leq d_f(x,z) + d_f(z,y)$
   for all $x,y,z$. First, we use that~$f$ is increasing
   and $d(x,y) \leq d(x,z) + d(z,y)$ (because~$d$ is a metric distance)
   to obtain
\begin{equation*}
   f(d(x,y)) \leq f(d(x,z) + d(z,y))
\end{equation*}
   Finally, given that~$f$ is concave and~$f(0) = 0$ we know that~$f$ is also
   subadditive and thus
\begin{equation*}
   f(d(x,z) + d(z,y)) \leq f(d(x,z)) + f(d(z,y))
\end{equation*}
\eolm\\
   The following lemma yields a quick way to determine whether a function
   is concave.

\textbf{Lemma 2}
   A function~$f$ that is twice differentiable on an interval~$I$ is concave
   on~$I$ if $f''(x) \leq 0$ for $x \in I$.

   The lemma can heuristically be understood as $f''(x) \leq 0$ implies that
   the rate of acceleration of~$f$ is slowing. Hence $f$~curves inward,
   implying it is concave.
   The lemma is part of standard calculus, and for a formal proof we refer to~\cite{hlp}.
   If $f$~is twice-differentiable, increasing, and satisfies $f''(x)\leq 0$ for
   $x>0$ with~$f(0)=0$ it is thus metric-preserving, and we will use this later.

\section{From correlations to distances}
   The first three properties of a metric distance are easily obtained when
   transforming one of the $A$ measures to a dissimilarity by a
   natural transformation such as $d(x,y) = 1 - A(x,y)$.
   However, the dissimilarity thus obtained does not guarantee the triangle
   inequality.  We show below why this is the case using generic principles rather
   than explicit calculations, and why transformations
   such as $d(x,y): x,y \rightarrow \sqrt{1-A(x,y)^2}$ and
   $d(x,y): x,y \rightarrow \sqrt{\oh(1 - A(x,y))}$
   do result in a metric distance.

   Currently two metric distances are known to derive from the triple $(P, S, C)$,
   namely the angle $\theta$ between vectors, and derived from it,
   $\sqrt{2-2 \cos(\theta)}$, which may be obtained as
   $\sqrt{2-2\;A(x,y)}$.
   For the angle $\theta$ the triangle inequality derives from Proposition
   \texttt{XI.20} of Euclid's The Elements and the fact that three vectors in a
   high-dimensional space can be embedded in three-dimensional space. It follows that
   $\arccos(A(x,y))$ yields a metric distance, where $A$ may be any of $P$, $S$, or $C$.
   It is known (e.g. \cite{sun::angular})
   that $\sqrt{2-2 \cos(\theta)}$ is equal to the Euclidean distance between the two
   unit-scaled object vectors~$x$ and~$y$. This follows from (using $\left\|x\right\| = 1$
   and $\left\|y\right\| = 1$)

\begin{align*}
\left\|x - y\right\|^2 &= \sum (x_i - y_i)^2\\
      \;    &= \left\|x\right\|^2 + \left\|y\right\|^2 - \left\|x\right\| \left\|y\right\| x \cdot y\\
            &= 2 - \cos(\theta)
\end{align*}

   It can additionally be
   observed using a trigonometric identity for $\sin(\oh\theta)$ (\cite{abrste}, page~72)
   that $\sqrt{2-2 \cos(\theta)}$ is equal to $\sqrt{2}\sin(\oh\theta)$ in the interval $[0,\pi]$
   and is seen to be concave on that interval by considering its second derivative.
   Hence $\sqrt{2}\;\sin(\oh\theta)$ is a metric-preserving function for
   the angular distance (but not metric-preserving in general).

   We formalise this finding and derive another class of metric
   distances derived from $P$, $S$, and $C$ whose members collate
   correlated and anti-correlated objects.
   The canonical representative of this class is the sine function $\sin$.
   In the lemma below we do not use generic metric-preserving
   functions, as stronger results can be obtained by utilising
   traits of the angular distance. However, the functions used share on certain
   intervals of interest the general traits of an important class of
   metric-preserving functions, namely being concave and increasing.

\textbf{Lemma 3}
i) A function $f$ of the angular distance that satisfies $f(0) = 0$, is
   defined on $[0, \pi]$, and is either
   a) increasing and concave on the interval $[0, \pi]$,
or b) increasing and concave on the interval $[0, \oh\pi]$  and
      satisfies $f(x) = f(\pi-x)$ ($f$ is symmetric around $\oh\pi$),
   is a metric preserving distance for the angular distance.
   In case b) this requires
   disregarding the directionality of vectors and collating a vector and
   its sign-reversed counterpart into a single object.

   Examples of such functions in case a) are\\
$f_1: x \rightarrow x$\\
$f_2: x \rightarrow \sin(\oh x)$

   Examples of such functions in case b) are\\
$f_3: x \rightarrow \oh\pi - |\oh\pi - x|$\\
$f_4: x \rightarrow \sin(x)$

   These lead to distances that can be computed, again using $A$ to denote any of $(P,S,C)$, as\\
$\;\;\;d_1: x,y \rightarrow f_1(A(x,y)) = \arccos(A(x,y))$\\
$\;\;\;d_2: x,y \rightarrow f_2(A(x,y)) = \sqrt{\oh(1-A(x,y))}$\\
(angular distance and correlation distance, respectively), and\\
$\;\;\;d_3: x,y \rightarrow f_3(A(x,y)) = \oh\pi - |\oh\pi - \arccos(A(x,y))|$\\
$\;\;\;d_4: x,y \rightarrow f_4(A(x,y)) = \sqrt{1-A(x,y)^2}$\\
(acute angular distance and absolute correlation distance, respectively).

   ii) A function $g$ of the angular distance that satisfies $g(0) = 0$ and
   is increasing and strictly convex on some interval $[0, \epsilon]$, where $\epsilon$
   is positive, yields a dissimilarity that violates the triangular inequality.
   An example of such a function is $g: x \rightarrow 1-\cos(x)$, or equivalently, $1-A(x,y)$.

\textbf{Proof}
   i) Name the three vectors $a$, $b$, and $c$ with angles $\alpha$, $\beta$, and $\gamma$ between the pairs
   $(b,c)$, $(a,c)$, and $(a,b)$
   respectively. In scenario a) we set out to prove that $f(\gamma) \leq f(\alpha) + f(\beta)$ and may use
   the inequality $\gamma \leq \alpha + \beta$ because the angular distance is a metric.
   In scenario a), if $\alpha+\beta \leq \pi$ we use subadditivity to deduce
   $f(\gamma) \leq f(\alpha+\beta) \leq f(\alpha) + f(\beta)$.
   In the other case it is easy to see that $f(\alpha) + f(\beta) \geq f(\pi)$, either
   by considering the concave function obtained by extending $f: x \rightarrow f(\pi)$ for $x > \pi$,
   or by explicit calculation. As $f(\pi)$ is the maximal value of $f$ in $[0,\pi]$ it follows that
   $f(\gamma) \leq f(\pi) \leq f(\alpha) + f(\beta)$.

   In scenario b) we may assume that $\alpha$ and $\beta$ are both smaller than $\oh\pi$ because of the following.
   By sign-reversing~$a$ we obtain vectors $-a, b, c$ and angles
   $\alpha, \pi-\beta, \pi-\gamma$.  This transform leaves the values of $f$ on the
   transformed angles invariant, and the triangular inequality can now be
   applied to $\alpha', \beta', \gamma'$ = $\alpha, \pi-\beta, \pi-\gamma$.
   Thus we may sign-reverse any of the three input vectors while preserving the inequality to be proven.
   By choosing which of $a$, $b$, or $c$ to flip we can always make sure that both
   $\alpha'$ and $\beta'$ are smaller than $\oh\pi$.  The inequality $f(\gamma) \leq f(\alpha) + f(\beta)$
   is the same as $f(\gamma') \leq f(\alpha') + f(\beta')$, where $\alpha'$,
   $\beta'$, $\gamma'$ are the angles corresponding with a triple of vectors $(a', b', c')$,
   allowing the use of the triangle inequality $\gamma' \leq \alpha' + \beta'$.
   If one of $\gamma'$ or $\pi-\gamma'$ is smaller than either of $\alpha'$ or
   $\beta'$ there is nothing to prove because $f$ is increasing on $[0, \oh\pi]$ and
   symmetric around $\oh\pi$.  If $\gamma'$ is bigger than $\oh\pi$, we observe that
   $\alpha' + \beta' \geq \gamma' \geq \pi-\gamma'$ and we can choose to work with
   $\gamma'' = \pi-\gamma'$ rather than $\gamma'$.  If $\gamma'$ is smaller than $\oh\pi$, we
   simply set $\gamma'' = \gamma'$.  This leaves us to prove $f(\gamma'') \leq f(\alpha') + f(\beta')$
   where $\gamma''$, $\alpha'$, and $\beta'$ are all smaller than
   $\oh\pi$, where $\gamma''$ is larger than both $\alpha'$ and $\beta'$, and where $\alpha'+\beta' \geq \gamma''$.
   The same reasoning as under a) now applies, restricted to the interval $[0,\oh \pi]$.

   ii) Pick vectors $a$, $b$ and $c$ lying in the cartesian plane, such that the angles
   satisfy $\gamma = \alpha + \beta$, $\gamma < \epsilon$.
   Then $g(\gamma) = g(\alpha+\beta) > g(\alpha) + g(\beta)$
   (by super-additivity of strictly convex functions with $f(0) \leq 0$).\eolm

\enlargethispage{3\baselineskip}

\section{Notes}
      For a distance~$d$ and a metric-preserving function~$f$ the distance~$d_f$ is
      ordinally equivalent with~$d$, that is, rankings of distances are preserved.
      The correlation distance $d_2$ is ordinally equivalent to the angular distance~$d_1$ and
      the acute angular distance~$d_3$ is equivalent to the absolute correlation distance~$d_4$.

      Further distances can be obtained by composition of concave functions;
      for example $f_5: x \rightarrow \sin(x)^p$, where $0<p\leq1$, also yields a distance.
      Such distances are again ordinally equivalent to the absolute correlation
      distance and preserve rankings of distances.

\section{Acknowledgments}
      The authors are grateful to Leopold Parts and Roberto \'Alvarez for
      critical reading and insightful comments.

\bibliographystyle{plos2009}
\bibliography{the}

\end{document}